\documentclass[prb,twocolumn,nopacs,nofootinbib]{revtex4-2}

\usepackage{amsmath}
\usepackage[dvipsnames]{xcolor}
\usepackage{nicefrac}
\usepackage{graphicx}
\usepackage{microtype}
\usepackage{physics}
\usepackage{mathrsfs}
\usepackage{bbold}
\usepackage{ dsfont }

\usepackage[bookmarks=true,colorlinks,linkcolor=blue,urlcolor=blue,citecolor=blue]{hyperref}

\usepackage[english]{babel}
\usepackage[applemac]{inputenc}
\usepackage[T1]{fontenc}

\newcommand{\beq}{\begin{equation}}
\newcommand{\eeq}{\end{equation}}
\newcommand{\beqa}{\begin{eqnarray}}
\newcommand{\eeqa}{\end{eqnarray}}

\usepackage{soul}
\usepackage[normalem]{ulem}

\graphicspath{{../}}

\begin{document}

\title{Ubiquity of the quantum boomerang effect in Hermitian Anderson-localized systems}
\author{Flavio Noronha$^{1}$ and Tommaso Macr\`{i}$^{2,1}$}
\affiliation{
$^{1}$Departamento de F\'{i}sica Te\'{o}rica e Experimental, Universidade Federal do Rio Grande do Norte, Campus Universit\'{a}rio, Lagoa Nova, Natal-RN 59078-970, Brazil\\
$^{2}$ITAMP, Harvard-Smithsonian Center for Astrophysics, Cambridge, Massachusetts 02138, USA
}

\date{\today}

\begin{abstract}
A particle with finite initial velocity in a disordered potential comes back and in average stops at the original location. This phenomenon dubbed `quantum boomerang effect' (QBE) has been recently observed in an experiment simulating the quantum kicked-rotor model [Phys. Rev. X {\bf 12}, 011035 (2022)].
We provide analytical arguments that support QBE in a wide class of disordered systems. Sufficient conditions to observe the {\it real-space} QBE effect are {\it (a)} Anderson localization, {\it (b)} the reality of the spectrum for the case of non-Hermitian systems,
{\it (c)} the ensemble of disorder realizations  $\{H\}$ be invariant under the application of $\mathcal{R\, T}$, and {\it (d)} the initial state is an {\it eigenvector} of $\mathcal{R\, T}$, where $\mathcal{R}$ is a reflection $x \rightarrow -x$ and $\mathcal{T}$ is the time-reversal operator.
The QBE can be observed in \textit{momentum-space} in systems with dynamical localization if conditions {\it (c)} and {\it (d)} are satisfied with respect to the operator $\mathcal{T}$ instead of $\mathcal{RT}$.
These conditions allow the observation of the QBE in {\it time-reversal} symmetry broken models, contrarily to what was expected from previous analyses of the effect, and in a large class of non-Hermitian models. We provide examples of QBE in lattice models with magnetic flux breaking time-reversal symmetry and in a model with electric field. 
Whereas the QBE straightforwardly applies to noninteracting many-body systems, 
we argue that a real-space (momentum-space) QBE is absent in weakly interacting bosonic systems due to the breaking of \textit{RT} (\textit{T}) symmetry. 
\end{abstract}

\maketitle
%\section{Introduction}\label{intro}
{\it Introduction. } 
The presence of disorder in a medium may lead to Anderson localization (AL) of quantum particles due to destructive interference~\cite{Anderson1958Absence}. 
AL has been experimentally observed in many platforms, including light~\cite{Chabanov2000,Schwartz2007}, ultrasound waves~\cite{Hu2008} and atomic matter~\cite{Julien2008,Manai2015,Billy2008,Jendrzejewski2012,Semeghini2015}. AL appears not only in Hermitian systems, but also in non-Hermitian models~\cite{HatanoNelson1996Localization,HatanoNelson1997Vortex,HatanoNelson1998Non-Hermitian,Efetov1997Directed,Feinberg1997Non-Hermitian,Feinberg99Non-Hermitian,Brouwer97Theory,Goldsheid98Distribution,Nelson98Non-Hermitian,HatanoNelson2016Non-Hermitian,Mudry98Random,Fukui98Breakdown,Yurkevich1999Delocalization,Longhi2015Robust,Zeng2017Anderson,McDonald2018Phase-Dependent,Gong2018,Hamazaki2019Non-Hermitian,Jiang2019Interplay,Zeng2020Topological,Kawabata2021Nounitary,Freilikher1994Effect,Beenakker1996Probability,Paasschens1996localization,Bruce1996Multiple,Longhi2019Topological,Tzortzakakis2020non-Hermitian,Huang2020Anderson}, which can be experimentally implemented with several platforms~\cite{Gong2018Topological,Scriffer2021Anderson,Weidemann2022Topological}.

One of the consequences of AL in the transport properties of a system is the quantum boomerang effect (QBE). It was theoretically shown that in the Anderson model the disorder-averaged center of mass (DACM) of a particle launched with a finite momentum $k_0$ would initially propagate ballistically, make a U-turn toward the origin after some time and stop at the initial position~\cite{Prat2019Quantum}. 
This phenomenon is different from the behavior expected for classical particles, where the center of mass would initially move away from the origin and saturate at a distance $\ell$ of the order of the mean free path.
%
%The QBE was expected to take place in general models with AL and time-reversal (\textit{T}) symmetry. 

In Ref.~\cite{Janarek2020Quantum} it was found that mean-field interactions in the Anderson model lead to the partial destruction of the QBE in the sense that the DACM stops after the U-turn, before reaching the origin. Recently, the presence of QBE was numerically shown in several \textit{T}-symmetric systems, including quasicrystals, models with disorder in the hoppings and in the quantum kicked rotor (QKR)~\cite{Tessieri2021Quantum}. 
The QKR presents AL and QBE in momentum space in the absence of interactions~\cite{Fishman1982,Tessieri2021Quantum}. When interactions are present, dynamical localization is destroyed~\cite{Cao2021}. 
%The noninteracting QKR can be an interesting model to investigate the relation between the QBE and the coherent backscattering and forward-scattering peaks~\cite{Lemarie2017,Muller2015}.
%The relation between the QBE and the coherent backscattering and forward-scattering peaks in the QKR has not been clearly understood yet~\cite{Lemarie2017,Muller2015}.
The existence of the QBE was confirmed in a very recent experimental implementation of the QKR~\cite{Sajjad2022}. The authors also showed the important role of time-reversal symmetry in that particular system, Floquet gauge and the initial state symmetry in supporting or disrupting the QBE. By using stochastic kicking in order to destroy AL, it was shown the breakdown of QBE.  
%\begin{figure}[b]
%\centering
%\includegraphics[width=\columnwidth]{Draft/figs/fig1.pdf}
%\caption{\textbf{QBE for \textit{RT}-symmetric ensemble of Hamiltonians and initial state}. ({\it a}) Hamiltonian $ H$ with onsite disorder $\{\varepsilon_j\}$, hopping $J_{j,q}$ and initial state $\psi_j = f_j\, e^{i k_0 j}$. Here we choose $f_j$ such that the initial state is an eigenvector of $\mathcal{R\,T}$ with $\mathcal{R\,T}\psi_j=-\psi_j$. ({\it b}) Hamiltonian $\tilde H=\mathcal{R\,T}  H \mathcal{R\,T}$ with disorder realization $\mathcal{R}\{\varepsilon_j\}$. }\label{fig:1}
%\end{figure} 
Moreover, all the previous results leading to the QBE were found in Hermitian \textit{T}-symmetric systems. Several questions arise as consequence of those findings, especially concerning the most general conditions to observe the QBE. 

In this work we provide analytical arguments for the presence of the QBE in a class of Hamiltonians much broader than the \textit{T}-symmetric ones, including both Hermitian and non-Hermitian models. We illustrate the validity of our analytical findings by means of numerical investigations showing the QBE in several models.

%\section{Conditions for the QBE}\label{conditions}
{\it Conditions for the QBE. } 
For compactness of notation, we consider here one-dimensional single-particle models. However, all the considerations below can be immediately generalized to an arbitrary number of spatial dimensions and many-body systems. 
%
%To investigate the QBE in real space one typically considers AL in an Hermitian $T$-symmetric Hamiltonian $H$ and a Gaussian initial wave packet
%$\psi_0 (x)=\langle x|\psi_0 \rangle={\mathcal N}\exp(-x^2/2\sigma^2+ik_0 x)$, where ${\mathcal N}$ is a normalization constant~\cite{Prat2019Quantum,Tessieri2021Quantum,Janarek2020Quantum}. 
We 
%generalize these assumptions and 
consider a Hamiltonian $H$ that may be either Hermitian or non-Hermitian. 
In the following, $\mathcal{T}$ is the time reversal operator and $\mathcal{R}$ is the reflection operator $\mathcal{R}:x\to -x$.
We will show that the QBE is expected to appear in real space if {\it (a)} the Hamiltonian presents AL, {\it (b)} all eigenenergies are real, {\it (c)} the ensemble $\{H\}$ of all disorder realizations of the model is \textit{RT} invariant, $\mathcal{RT}\{H\}(\mathcal{RT})^{-1}= \{H\}$,
and {\it (d)} the initial state is an eigenstate of $\mathcal{RT}$, $\mathcal{RT} |\psi_0 \rangle=\pm |\psi_0 \rangle$. 
%\begin{equation}
%\mathcal{RT} |\psi_0 \rangle=\pm |\psi_0 \rangle.\label{PTpsi}   
%\end{equation}
%Conditions {\it (c)} and {\it (d)} are illustrated in Fig.~\ref{fig:1}.

Without loss of generality we assume that the center of mass of the initial wave packet is positioned at the origin. 
We can expand $\ket{\psi_0}=\sum_n c_n \ket{\phi_n}$ in terms of the eigenvectors of the Hamiltonian, $H\ket{\phi_n}=\epsilon_n \ket{\phi_n}$.
Using condition {\it (b)} we find that,
at an arbitrary time $t$, the center of mass is given by %$\langle x(t)\rangle = \int_{-\infty}^{+\infty}\, x \, \left|\psi (x,t)\right|^2\, dx$, 
\beq 
\langle x(t)\rangle=\sum_{n,m} c_n c_m^*\,\textrm{exp}[-i(\epsilon_n - \epsilon_m)t]\bra{\phi_m}X \ket{\phi_n}, \label{expect}
\eeq 
where $X$ is the position operator.
%\begin{equation}
%    \langle x(t)\rangle = \int_{-\infty}^{+\infty}\, dx \, \left|\psi (x,t)\right|^2\, x,\label{expect}
%\end{equation}
%where we can write 
%\begin{eqnarray}\nonumber
%\left|\psi (x,t)\right|^2 &=& %\left|\textrm{e}^{-i Ht}\psi_0 (x)\right|^2=
%\bra{\psi_{0} }\textrm{exp}(+i H^\dagger t)\ketbra{x}{x}\textrm{exp}(-i Ht)\ket{\psi_0 }\\
%\sum_{n,m} \phi_{m}(x)^{*} \phi_{n}(x)\,\textrm{exp}[-i(\epsilon_n - \epsilon_m^*)t]\nonumber\\
%& & \times \braket{\psi_{0} }{\phi_m}\braket{\phi_n}{\psi_{0} }.\label{psixt}
%\end{eqnarray}
%Here we have defined $\phi_n(x)=\bra{x} \ket{\phi_n}$ in terms of the eigenvectors $\ket{\phi_n}$ of the Hamiltonian $ H$. 
%which satisfy $ H\ket{\phi_n}=\epsilon_n \ket{\phi_n}$ and $\bra{\phi_n} H^\dagger=\epsilon_n^*\bra{\phi_n}$.
%
Using condition {\it (a)} we have, after averaging over many disorder realizations and taking the limit $t\to+\infty$, the diagonal ensemble~\cite{Prat2019Quantum,Sajjad2022,footeq2}
\begin{eqnarray}
\overline{\langle x(+\infty)\rangle}
&=&\sum_n\overline{\left|c_n\right|^2\bra{\phi_n }X\ket{\phi_n}},\label{diagems}
\end{eqnarray}
%
%\begin{eqnarray}
%\overline{\left|\psi (x,+\infty)\right|^2}&=&\sum_n\overline{\left|\phi_{n}(x)\right|^2\left|\braket{\psi_0 }{\phi_n}\right|^2},\label{diagems}
%\end{eqnarray}
where the overline $\overline{(\cdots)}$ denotes average over the disorder realizations.
%The above relation is not necessarily valid without taking disorder average.
An equivalent expression is found when one takes the limit $t\to-\infty$ and hence %we conclude that $\overline{\left|\psi (x,+\infty)\right|^2}=\overline{\left|\psi (x,-\infty)\right|^2}$.
%\begin{eqnarray}
%\overline{\left|\psi (x,+\infty)\right|^2}&=&\overline{\left|\psi (x,-\infty)\right|^2}.\label{diag1}
%\end{eqnarray}
%The average center of mass is given by
%\begin{equation}
 %   \overline{\langle\hat{x}(t)\rangle}=\int_{-\infty}^{+\infty}dx \overline{\left|\psi (x,t)\right|^2} x.
%\end{equation}
%Therefore,
\begin{eqnarray}
\overline{\langle x(+\infty)\rangle} &=& \overline{\langle x(-\infty)\rangle}.\label{eqdiag}
\end{eqnarray}
For each disorder realization $H$ we define its \textit{RT} counterpart $\tilde{H}=\mathcal{RT} H(\mathcal{RT})^{-1}$.
%Notice that this does not necessarily imply that $\tilde{H}={\mathcal{RT}} H{\mathcal{TR}}=H$, but only that the disorder average of an observable ${\mathcal{O}}$ computed on $H$ and $\tilde{H}$ give the same result, i.e.~$\overline{\langle {\mathcal O} \rangle_H}=\overline{\langle {\mathcal O} \rangle_{\tilde{H}}}$.
The center of mass of a state evolved under the disorder realization $H$ satisfies
\begin{eqnarray}
\langle x(t)\rangle_{ H}%&=&\int dx \left|\psi (x,t)\right|^2 x
%&=&\bra{\psi_0 }  
%\textrm{exp}(i H^\dagger t) ~X~ \textrm{exp}(-i H t)
%\ket{\psi_0 }\nonumber\\
%
%
&=&(\pm\bra{\psi_0 })  [\mathcal{RT}
\textrm{exp}(i H^\dagger t)(\mathcal{RT})^{-1}][\mathcal{RT} X(\mathcal{RT})^{-1}]\nonumber\\
& &\times [\mathcal{RT} \textrm{exp}(-i H t)(\mathcal{RT})^{-1}]
(\pm\ket{\psi_0 })\nonumber\\
%
%&=&\bra{\psi_0 }  \textrm{exp}(-i\tilde{H}^\dagger t) (-X) \textrm{exp}(i\tilde{H} t) \ket{\psi_0 }\nonumber\\
%
&=& -\langle x(-t)\rangle_{\tilde{H}},\label{QPT}
\end{eqnarray}
where we have used condition \textit{(d)}. % $\mathcal{PT} \ket{\psi_0 }=\ket{\psi_0 }$. 
%
%We want to prove that for $ H=\hat{p}^2/2m + V(\hat{x})$ and $\overline{V(x)}=0$ and $\overline{V(x)V(x')}=W\delta(x-x')$, we have $\overline{\langle \hat{x} \rangle_{\tilde{H}}}(t)=\overline{\langle \hat{x} \rangle_{ H}}(t)$, where $\tilde{H}=\hat{p}^2/2m +V(-\hat{x})$. When averaging over the disorder we take the average over several (in principle infinite) realizations. Then $\overline{\langle \hat{x}\rangle}=\sum_j\langle\hat{x}\rangle_j$, where
%\begin{eqnarray}\nonumber
%\langle\hat{x}\rangle_j &=& \int  dx %\left|\psi(x,t)\right|^2 x\\
%&=& \bra{\psi_0}\textrm{e}^{i\left(\frac{p^2}{2m}+V_i(\hat{x})\right)t}\hat{x}\textrm{e}^{-i\left(\frac{p^2}{2m}+V_i(\hat{x})\right)t}\ket{\psi_0},
%\end{eqnarray}
%$\exists$ $j_1$ and $j_2:$ $V_{j_1}(-x)=V_{j_2}(x)$. 
%
%If the model is such that for each disorder realization $ H$ we have that $\tilde{H}=\hat{\mathcal{P}} H\hat{\mathcal{P}}$ is also a disorder realization,
Now we use condition \textit{(c)}, which is equivalent to say that for each disorder realization $H$ its \textit{RT} counterpart $\tilde H$ is also a disorder realization of the same model. Therefore $\overline{\langle x(t)\rangle} = -\overline{\langle x(-t)\rangle}$  %$\overline{\langle\hat{x}\rangle_{\tilde{H}}(t)}=\overline{\langle\hat{x}\rangle_{ H}(t)}$. Then we have
%\begin{eqnarray}
%\overline{\langle x(t)\rangle} &=& -\overline{\langle x(-t)\rangle}
%\end{eqnarray}
and, in particular, 
\begin{eqnarray}
\overline{\langle x(+\infty)\rangle} &=& -\overline{\langle x(-\infty)\rangle}.\label{eqder1}
\end{eqnarray}
From Eqs.~(\ref{eqdiag}) and (\ref{eqder1})  %$\overline{\langle x\rangle(+\infty)} =\overline{\langle x\rangle(-\infty)}$. 
we have $\overline{\langle x(+\infty)\rangle}=0$,
%\begin{eqnarray}
%\overline{\langle x(+\infty)\rangle}=0\label{vanish}
%\end{eqnarray}
which guarantees that the QBE occurs.

%Our arguments can be easily generalized to any higher dimension $d>1$. In this case each of the coordinates $x_1,\cdots,x_d$ presents the QBE if Eq.~(\ref{PTpsi}) is satisfied and for each realization $H$ there is a realization ${\mathcal PT} H {\mathcal PT}$. However, in order to observe the QBE in a single coordinate, say $x_1$, it is enough to consider the reflection operator ${\mathcal P}:(x_1,x_2,\cdots,x_d)\rightarrow (-x_1,x_2,\cdots,x_d)$ instead of ${\mathcal P}$. Then the initial state is required to satisfy ${\mathcal PT} \ket{\psi_0}=\pm \ket{\psi_0}$ and for each realization $H$ one needs a realization ${\mathcal PT} H {\mathcal PT}$.

In higher dimensions, without loss of generality, the initial momentum is chosen to be aligned along the $X$ direction 
%In higher dimensions the coordinate operator $X$ is defined along the direction of the initial momentum
and $\mathcal{R}$ is the reflection with respect to $X$.
In cases where conditions \textit{(c)} and \textit{(d)} are not satisfied with respect to the operator $\mathcal{RT}$ it is still possible to guarantee the QBE if 
there is some unitary operator $\mathcal{U}$ that commutes with $X$ and causes conditions \textit{(c)} and \textit{(d)} to be satisfied with respect to $\mathcal{URT}$ [see the Supplemental Material (SM)~\cite{Supplement} for details on the derivation].
%these conditions are satisfied with respect to an operator $\mathcal{URT}$, where $\mathcal{U}$ is any unitary operator that commutes with $X$ 
%~\cite{Supplement}.
In models that present localization in momentum space, e.g.~the QKR, the QBE can appear in $\langle p(t)\rangle$~\cite{Tessieri2021Quantum,Sajjad2022}. The demonstration of this effect in momentum space follows the arguments that we have shown above but considers only the operator ${\mathcal T}$ instead of $\mathcal{RT}$ in conditions \textit{(c)} and \textit{(d)} (see SM~\cite{Supplement}). 
In the following we 
%use the Schr{\"o}dinger equation and 
show numerically the QBE in several Hermitian models in which the presented analytical arguments apply. 
In Ref.~\cite{nonHermitian} we confirm numerically the QBE in several non-Hermitian models and show its main features in those systems.

\begin{figure}[b]
\centering
\includegraphics[width=\columnwidth]{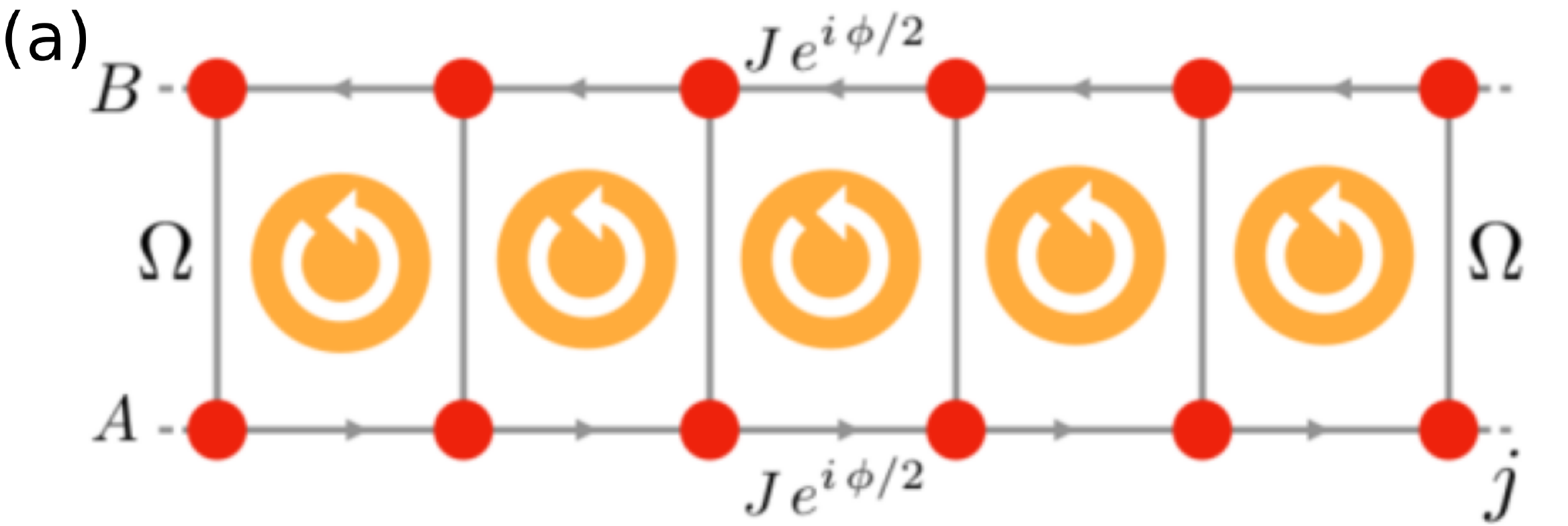}
\includegraphics[width=\columnwidth]{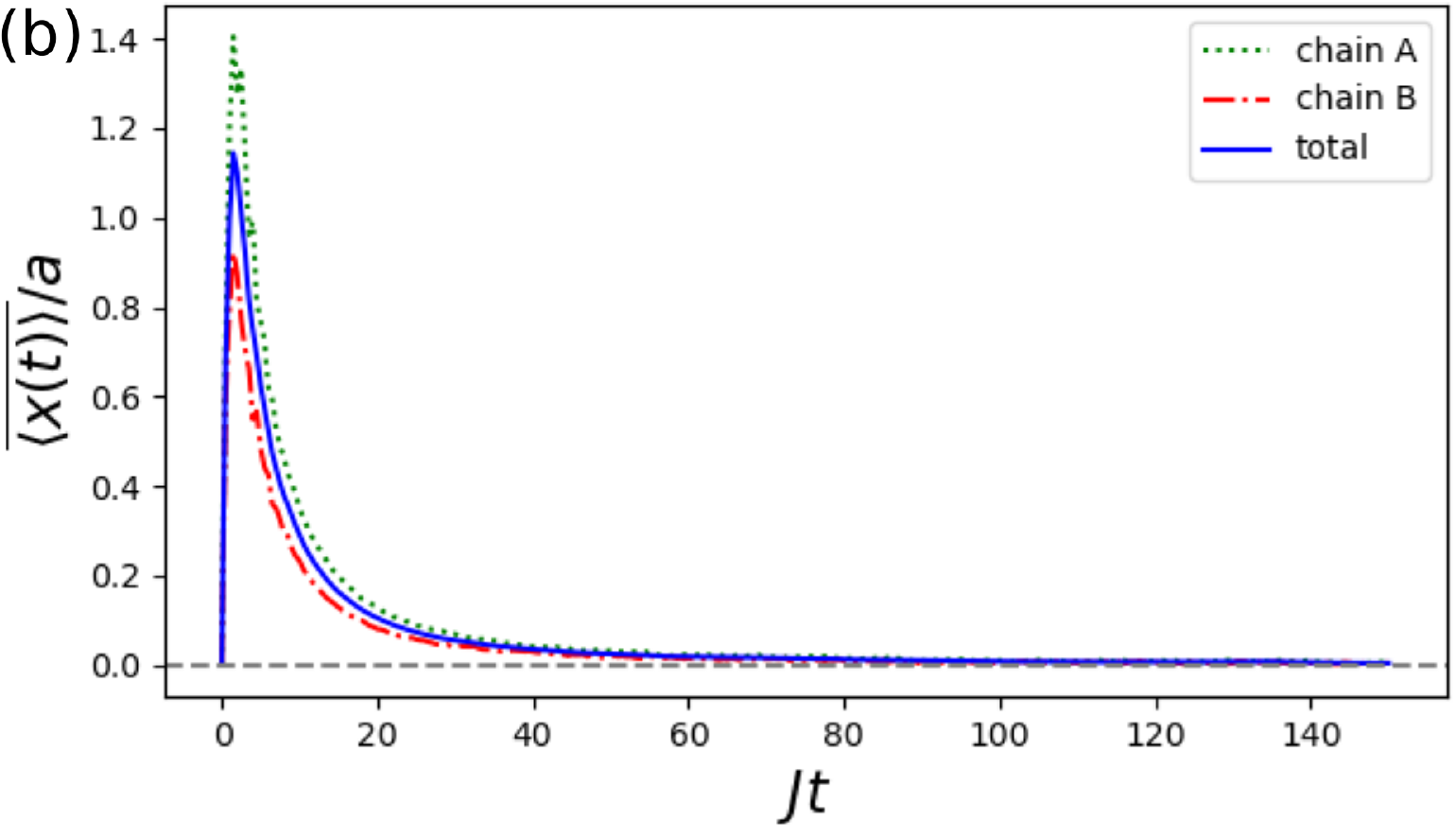}
\caption{\textbf{QBE with broken \textit{T} symmetry: The Harper-Hofstadter ladder}. 
({\it a}) Two-leg ladder model with magnetic flux $\phi$ per plaquette, intra-chain complex hopping $J\, e^{\pm i\phi/2}$ and inter-chain coupling $\Omega$.
({\it b}) Disorder averaged center of mass in chain A (green dotted), chain B (red dash-dotted) and averaged in the whole system (blue solid). We set $\Omega/J=2$, $W/J=6$, $\phi=(\pi/2)(\sqrt{5}-1)/2$, $N=4\times 10^2$, $n_d=5\times10^5$ and initial state $\psi_{1+}$ with $\sigma/a=10$ and $k_0a=1.4$.}\label{figcreutz}
\end{figure}

%\section{QBE in models with magnetic field}\label{magnetic}
{\it QBE in models with magnetic field. } 
%\textit{QBE in models with magnetic field.}
In this section we show the QBE in two different models that break \textit{T} symmetry by means of a magnetic field, the Harper-Hofstadter ladder model and the 2D Harper model. 
%\subsection{Harper-Hofstadter ladder model}
In order to demonstrate the QBE in a minimal model that breaks \textit{T} symmetry we consider first the Harper-Hofstadter ladder model~\cite{Harper1955General,Hofstadter1976,Streda1982,WuCreutz2022}, illustrated in Fig.~\ref{figcreutz}(a) and described by the Hamiltonian
\beqa
%H&=&H_0+H_1\nonumber \\
H&=&-J\sum_i[\textrm{e}^{i\phi/2}(b_{i}^\dagger b_{i+1}+a_{i+1}^\dagger a_{i})+H.c.]\nonumber\\
& &-\Omega\sum_i [a_i^\dagger b_i+H.c.]
+\sum_i \epsilon_{a,i}a_i^\dagger a_i+\epsilon_{b,i}b_i^\dagger b_i, \quad  \label{creutzm}%\nonumber\\ & &\label{creutzm}
\eeqa  
where $i=1,\cdots,N$, $N$ is the number of sites in each of the chains $A$ and $B$, and $a_i^\dagger$ ($b_i^\dagger$) creates a particle on site $i$ of chain $A$ ($B$). $J\textrm{e}^{\pm i\phi/2}$ characterize the intra-chain hoppings, 
where $J,\phi\in \mathds{R}$, while $\Omega\in \mathds{R}$ is the inter-chain hopping amplitude. 
We consider open boundary conditions in each chain. The onsite potentials $\epsilon_{a,i},\epsilon_{b,i}$ are uncorrelated random numbers sampled from a uniform distribution over $[-W/2,W/2]$. The model presents Anderson localization due to the disorder. 

Here we define the reflection operator as $\mathcal{R}: (a_i, b_i)\rightarrow (a_{-i}, b_{-i})$ and the time-reversal operator $\mathcal{T}=\mathcal{K}$ is the complex conjugation. Decomposing the Hamiltonian $H=H_0 +H_1$ into a hopping term $H_0 $ and a local potential term $H_1$, one can check that $H_0 $ breaks time-reversal symmetry due to the complex hoppings $J\textrm{e}^{\pm i\phi/2}$. This symmetry breaking is related to a magnetic flux through each plaquette of the ladder, which is proportional to the phase $\phi$ acquired along the loop around each plaquette. The hopping term satisfies $\mathcal{R} \mathcal{T}H_0 (\mathcal{R} \mathcal{T})^{-1} =H_0$.
%\beqa
% \mathcal{R} \mathcal{T}H_0 \mathcal{T}\mathcal{R} &=&H_0 .\label{symH}
%\eeqa 
The ensemble of disorder realizations is \textit{RT} invariant, $\mathcal{R} \mathcal{T}\{H\}(\mathcal{R} \mathcal{T})^{-1} =\{H\}$. Therefore the QBE is expected to appear if condition \textit{(d)} is satisfied.%we consider as initial wave packet an eigenstate of $\mathcal{R} \mathcal{T}$.

A wave packet of the system may be written as a spinor with components in chains $A$ and $B$, in the form $\psi(x_j)=\big(\psi^{(a)}(x_j),\psi^{(b)}(x_j)\big)$. We define four wave packets
\beqa 
\psi_{0\pm}(x_j)&=&\mathcal{N}_0\,\exp(-x_j^2/2\sigma^2+ik_0x_j) (1,\pm 1),\nonumber\\
\psi_{1\pm}(x_j)&=&\mathcal{N}_1\,x_j\,\exp(-x_j^2/2\sigma^2+ik_0x_j) (1,\pm 1), \quad %\nonumber 
\eeqa 
where $\mathcal{N}_0$ and $\mathcal{N}_1$ are normalization factors and $x_j$ is the position of site $j$ (for simplicity we consider an unitary lattice parameter $a=1$). 
These wave packets satisfy
$\mathcal{R} \mathcal{T}\psi_{0\pm}=+ \psi_{0\pm}$, $\mathcal{R} \mathcal{T}\psi_{1\pm}=- \psi_{1\pm}$.
%\beqa 
%\mathcal{R} \mathcal{T}\psi_{0\pm}&=&+ \psi_{0\pm},\nonumber\\
%\mathcal{R} \mathcal{T}\psi_{1\pm}&=&- \psi_{1\pm}.\label{relationspsi}
%\eeqa 
%
As a consequence, the QBE appears in the total center of mass $\overline{\langle x(t) \rangle}=\sum_i x_i\big[\overline{|\psi^{(a)}(x_i,t)|^2}+\overline{|\psi^{(b)}(x_i,t)|^2}\big]$ using any of the four initial wave packets above [see in Fig.~\ref{figcreutz}(b) the QBE using $\psi_{1+}$].
This shows that the initial wave function does not need to be invariant under $\mathcal{RT}$, but it is enough to be its eigenstate.
Moreover, the QBE is present in each chain individually through $\overline{\langle x(t) \rangle}_l=\big[\sum_i x_i\overline{|\psi^{(l)}(x_i,t)|^2}\big]/\sum_i \overline{|\psi^{(l)}(x_i,t)|^2}$, $l=a,b$.
The validity of the QBE in each chain can be checked analytically through Eq.~(\ref{QPT}) using $\Pi_l\, X\, \Pi_l$ instead of $X$, where $\Pi_l$ is the projection operator on chain $l=a,b$. Aditional data for the Harper-Hofstadter ladder model is available in the SM~\cite{Supplement}.

The Harper-Hofstadter ladder model can also be interpreted as composed of spin-1/2 particles on a chain, and in Eq.~(\ref{creutzm}) $a_i^\dagger$ ($b_i^\dagger$) creates at site $i$ a spin up (down) fermion. In this case, $\mathcal{T}=\mathcal{SK}$ takes the complex conjugate~($\mathcal{K}$) and flips the spin indices ($\mathcal{S}=\sigma_x$). %, while $\mathcal{P}=\mathcal{R}$. 
Conditions \textit{(c)}-\textit{(d)} are not met with respect to the operator $\mathcal{RT}=\sigma_x\mathcal{RK}$.
Therefore we choose $\mathcal{U}=\mathcal{S}^{-1}$ so the operator $\mathcal{URT}=\mathcal{R}\mathcal{K}$ acts in the same way that it acted in the previous interpretation of the Harper-Hofstadter ladder. Therefore the ensemble of all disorder realizations satisfies $\mathcal{URT} \{H\} (\mathcal{URT})^{-1}=\{H\}$ and $\psi_{0\pm}$, $\psi_{1\pm}$ defined above are eigenvectors of $\mathcal{URT}$. This leads to the QBE, illustrating that our analytical arguments also apply in the case of particles with spin.

%Notice, however, that $\varphi_{\pm}(x_i)=\alpha \psi_{0\pm}(x_i)+\beta\psi_{1\mp}(x_i)$ is not an eigenstate of $\mathcal{R} \mathcal{T}$ and using this function as initial wave packet leads to the breakdown of the QBE.

%\subsection{Harper Model}
%{\it QBE in models with magnetic field. --}
%In the literature of the QBE, all considered models that display this effect are time-reversal invariant~\cite{Prat2019Quantum,Tessieri2021Quantum,Sajjad2022}.
Here, in order to further investigate the importance of conditions \textit{(c)}-\textit{(d)}, we consider the presence of disorder in the Harper model of a 2D lattice with an external magnetic field, given by the Hamiltonian~\cite{Harper1955,Hofstadter1976}
\beqa
%H&=&H_0+H_1\nonumber \\
H&=&-J\sum_{j,l}[\textrm{e}^{-i 2\pi\alpha l}c_{j+1,l}^\dagger \, c_{j,l}+c_{j,l+1}^\dagger \, c_{j,l}+H.c.]\nonumber\\
& &+\sum_{j,l} \epsilon_{j,l}\, c_{j,l}^\dagger \, c_{j,l},  \label{HarperH}%\nonumber\\ & &\label{creutzm}
\eeqa  
where $j=1,...,N_x$ ($l=1,...,N_y$) characterizes the $x$ ($y$) coordinate of the system with lattice parameter $a=1$ and $c_{j,l}^\dagger$ creates a particle on site $(j,l)$. The
complex coefficients $J\textrm{e}^{\mp i 2\pi\alpha l}$ define the hoppings in the horizontal direction and $J\in \mathds{R}$ is the hopping in the vertical direction. $\alpha$ is proportional to the magnetic flux in each plaquette.
%of the lattice and the system is invariant under $\alpha\to \alpha+1$. 
%
\begin{figure}[t]
\centering
\includegraphics[width=\columnwidth]{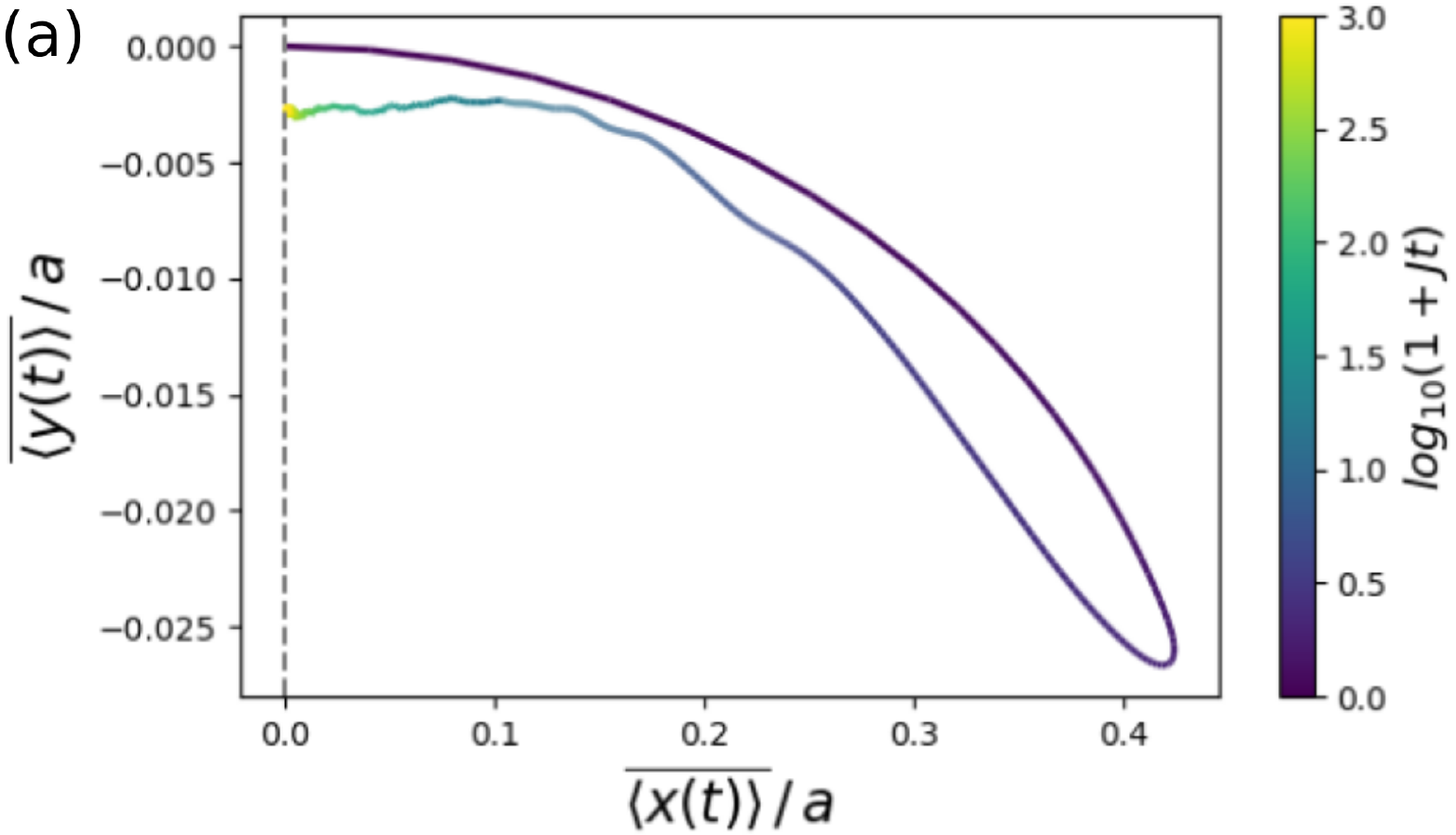}
\includegraphics[width=\columnwidth]{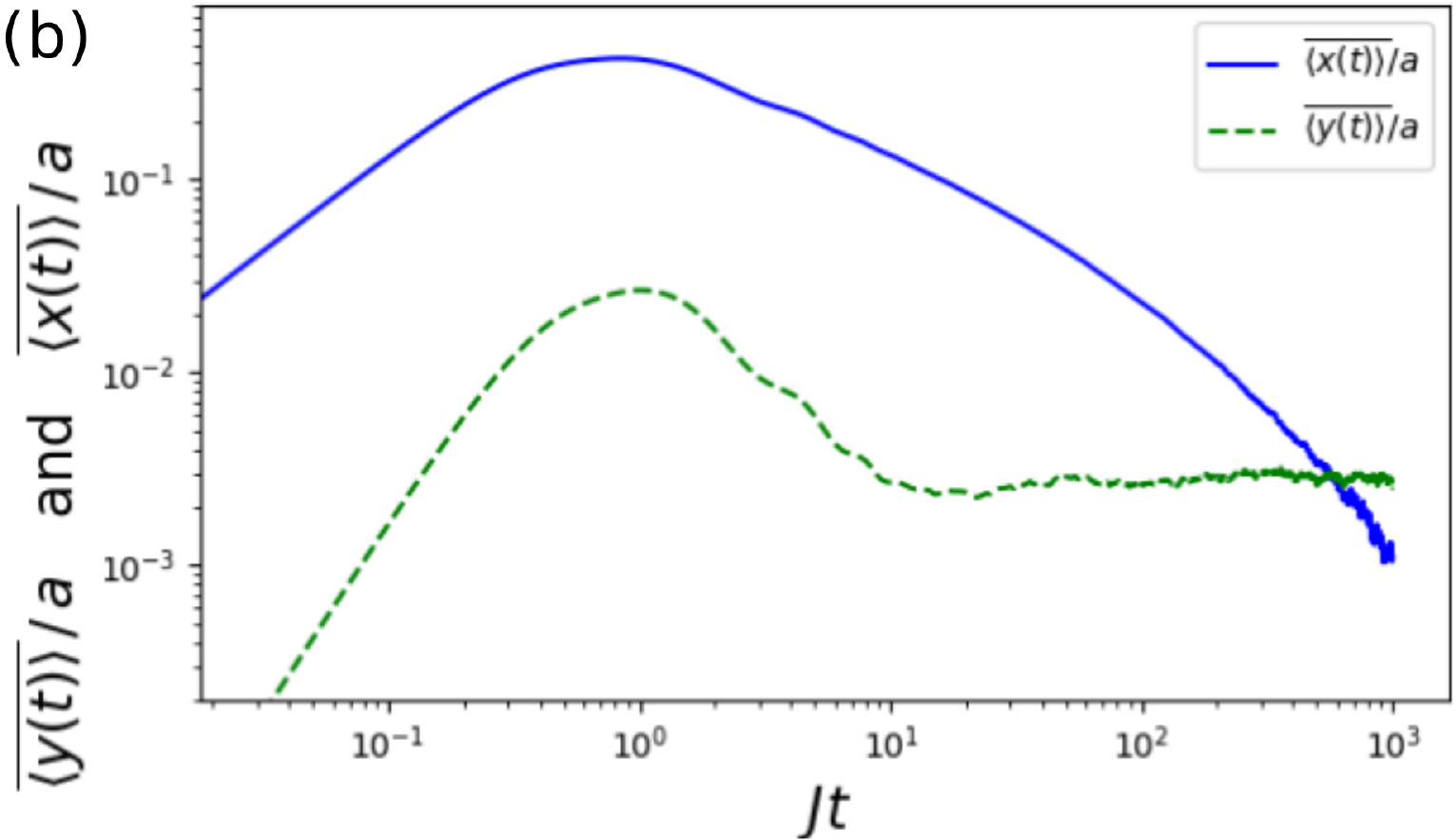}
\caption{\textbf{QBE with broken \textit{T} symmetry: The Harper model}. 
(a) Trajectory of $\overline{\langle x(t)\rangle}$ $\times$ $\overline{\langle y(t)\rangle}$ presenting the full (partial) boomerang effect in the direction parallel (perpendicular) to the initial momentum $\boldsymbol{k}_0=k_0 \hat{x}$, i.e.~$\overline{\langle x(+\infty)\rangle}=0$ ($\overline{\langle y(+\infty)\rangle}\neq 0$). The color bar indicates the time propagation in the interval $Jt\in [0,1000]$.
(b) In the blue solid line we show that $\overline{\langle x(t)\rangle}$ decreases and tends to vanish and in the green dashed line we show that $\overline{\langle y(t)\rangle}$ remains finite at long times. We set $W/J=10$, $\alpha=0.02$, $n_d=7\times 10^5$ disorder realizations, $\sigma/a=10$, $k_0a=\pi/2$ and $N_x=N_y=190$.}\label{HarperQBE}
\end{figure}
We consider open boundary conditions. The onsite potentials $\epsilon_{j,l}$ are uncorrelated random numbers sampled from a uniform distribution over $[-W/2,W/2]$. 
Because of the disorder this model presents AL~\cite{Abrahams1979,Lee1985,Kramer1993,Wegner1989,MacKinnon1983,Muller1997}.

Here we define the reflection operators as $\mathcal{R}_x: c_{j,l}\rightarrow c_{-j,l}$, $\mathcal{R}_y: c_{j,l}\rightarrow c_{j,-l}$ and the time-reversal operator $\mathcal{T}=\mathcal{K}$ is the complex conjugation. Decomposing the Hamiltonian $H=H_0 +H_1$ into a hopping term $H_0 $ and a local potential term $H_1$, one can check that $H_0 $ breaks time-reversal symmetry due to the complex hoppings. Though $H_0$ is not \textit{T} symmetric, it satisfies $\mathcal{R}_x \mathcal{T}H_0 (\mathcal{R}_x \mathcal{T})^{-1}=\mathcal{R}_y \mathcal{T}H_0 (\mathcal{R}_y \mathcal{T})^{-1} =H_0$.
%\beqa
% \mathcal{R} \mathcal{T}H_0 \mathcal{T}\mathcal{R} &=&H_0 .\label{symH}
%\eeqa 
The ensemble of disorder realizations is \textit{RT} invariant, $\mathcal{R}_x \mathcal{T}\{H\}(\mathcal{R}_x \mathcal{T})^{-1}=\mathcal{R}_y \mathcal{T}\{H\}(\mathcal{R}_y \mathcal{T})^{-1} =\{H\}$. Therefore the QBE is expected to appear if condition \textit{(d)} is satisfied.%we consider as initial wave packet an eigenstate of $\mathcal{R} \mathcal{T}$.

We initialize the system in a Gaussian wave packet, $\psi_{0}(\boldsymbol{r}_{j,l})=\mathcal{N}_0\,\textrm{exp} (-r_{j,l}^2/2\sigma^2+i \boldsymbol{k}_0\cdot \boldsymbol{r}_{j,l})$, where $\boldsymbol{r}_{j,l}$ is the position of site $(j,l)$. Without loss of generality we consider $\boldsymbol{k}_0=k_0 \hat{x}$. This wave function satisfies $\mathcal{R}_x\mathcal{T} \psi_{0}(\boldsymbol{r}_{j,l})=\psi_{0}(\boldsymbol{r}_{j,l})$ and hence the QBE is expected to take place in the direction of the initial momentum, i.e.~in $\overline{\langle x(t)\rangle}$. In the perpendicular direction we have $\mathcal{R}_y\mathcal{T} \psi_{0}(\boldsymbol{r}_{j,l})=\psi_{0}(\boldsymbol{r}_{j,l})^*\neq \psi_{0}(\boldsymbol{r}_{j,l})$ and our analytical arguments do not guarantee that the QBE will take place in $\overline{\langle y(t)\rangle}$. We check numerically that the QBE appears in the $x$ direction but is broken in the $y$ direction; after the U-turn $\overline{\langle y(t)\rangle}$ does not reach the origin (see Fig.~\ref{HarperQBE}). %We observe this phenomenon using different values of parameters like $k_0, \alpha$ and $W$.
This confirms the presence of the QBE in \textit{T}-broken models and illustrates the importance of conditions \textit{(c)}-\textit{(d)}.

\begin{figure}[t]
\centering
\includegraphics[width=\columnwidth]{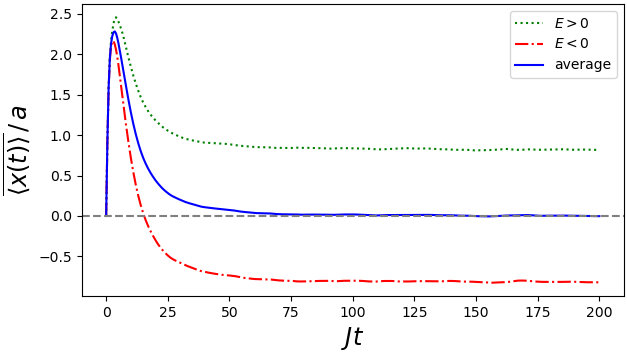}
\caption{\textbf{QBE in the Anderson model with electric field}. 
Using $W/J=3$ and $|E|/J=0.1$, we show the disorder-averaged center of mass for the case with $E>0$ ($E<0$) in green dotted (red dot-dashed) line. The average of these two cases is shown in blue solid line. In these data we considered a Gaussian initial state and used $\sigma/a=10$, $k_0a=1.4$, $N=4\times 10^2$ and $n_d=5\times10^4$.}\label{FElectric}
\end{figure}

%\section{Anderson model with electric field}\label{electric}
{\it Anderson model with electric field. }
Another interesting case is the $1$D Anderson model in the presence of an external electric field~$E$. 
%To satisfy $\mathcal{RT}\{H\}(\mathcal{RT})^{-1}=\{H\}$ one can consider the union of the ensemble of realizations with $+E$ with that with $-E$, i.e.~$\{H\}=\{H(+E)\} \cup \{H(-E)\}$ (see~\cite{Supplement}).
%
%We verify that the QBE also appears in the presence of an external electric field $E$.
%To guarantee $\mathcal{RT}\{H\}(\mathcal{RT})^{-1}=\{H\}$ we consider the union of the ensemble of realizations with $+E$ with that with $-E$, i.e.~$\{H\}=\{H(+E)\} \cup \{H(-E)\}$.
%The $1$D Anderson model in an electric field $E\in \mathds{R}$ reads
The model reads
\begin{eqnarray}\label{ElectricH}
H=\sum_{j} \left[-J c_{j+1}^\dagger c_j - J c_{j}^\dagger c_{j+1} +(\epsilon_j -jE)c_j^\dagger c_j \right],
\end{eqnarray}
where $\epsilon_j$ are sampled from a uniform distribution over $[-W/2,W/2]$.
The ensemble of disorder realizations with field $E$ satisfies $\mathcal{RT}\{H(E)\}(\mathcal{RT})^{-1}=\{H(-E)\}$ and hence the QBE is not observed when averaging $\langle x(t)\rangle$ over $\{H(E)\}$ [see green dotted (red dot-dashed) line in Fig.~\ref{FElectric} for $E>0$ ($E<0$)].
To guarantee $\mathcal{RT}\{H\}(\mathcal{RT})^{-1}=\{H\}$ we consider the union of the ensemble of disorder realizations with field $+E$ with the realizations with $-E$, i.e.~$\{H\}=\{H(+E)\} \cup \{H(-E)\}$.
Figure~\ref{FElectric} shows in blue solid line the presence of the QBE in this case. 
%Notice that the dynamics of the DACM obtained here is not identical to the one of the Anderson model in the absence of electric field. 
Notice that this is not equivalent to take the average of the Hamiltonians with $E>0$ with those with $E<0$ and obtain the Anderson model in the absence of $E$.
These results further illustrate the importance of condition \textit{(c)}.
%The average of $\langle x(t)\rangle$ over only $\{H(+E)\}$ or $\{H(-E)\}$ leads to the absence of the boomerang (also shown in Fig.~\ref{FElectric}). 

%\section{Many-body systems}\label{manyb}
{\it Many-body systems. }
Non-interacting many-particle systems satisfying the conditions mentioned in the analytical arguments are expected to display the QBE. 
In fact, it is straightforward to prove that for an initial $N$-particle bosonic (B) or fermionic (F) state $\psi_{B,F}(x_1,\dots,x_N)=\left<x_1,\dots,x_N|\chi_1,\dots,\chi_N\right>_{B,F}$ one has $\langle  X(t)\rangle  = \sum_{i} \left<\chi_i(t)| X_i|\chi_i(t)\right>=\sum_i\langle x_i(t)\rangle$, where $X=\sum_{i} X_i$, $X_i$ is the position operator corresponding to the $i$-th particle, $\langle x_i(t)\rangle$ is its center of mass and $\chi_i$ is the $i$-th orbital, $i=1,\dots,N$. 
Therefore the QBE appears in $\overline{\langle  x_i(t)\rangle}$ and hence in $\overline{\langle  X(t)\rangle}$ averaging over disorder realizations. We  also notice that, if there is a sufficiently large number $N$ of particles far from each other, each of them feels a different local disorder in its vicinity and the summation $\sum_i\langle x_i(t)\rangle$ plays the role of average over disorder realizations. Therefore, for a single disorder realization the QBE is also expected to appear in the average center of mass of the system $\langle X(t)\rangle/N$.
A similar argument holds for the QBE in momentum space. 
Therefore, in the case of non-interacting many particles we expect the QBE to appear even in the presence of electric or magnetic fields if conditions \textit{(a)}-\textit{(d)} are met.

\begin{figure}[b]
\includegraphics[width=0.95\columnwidth]{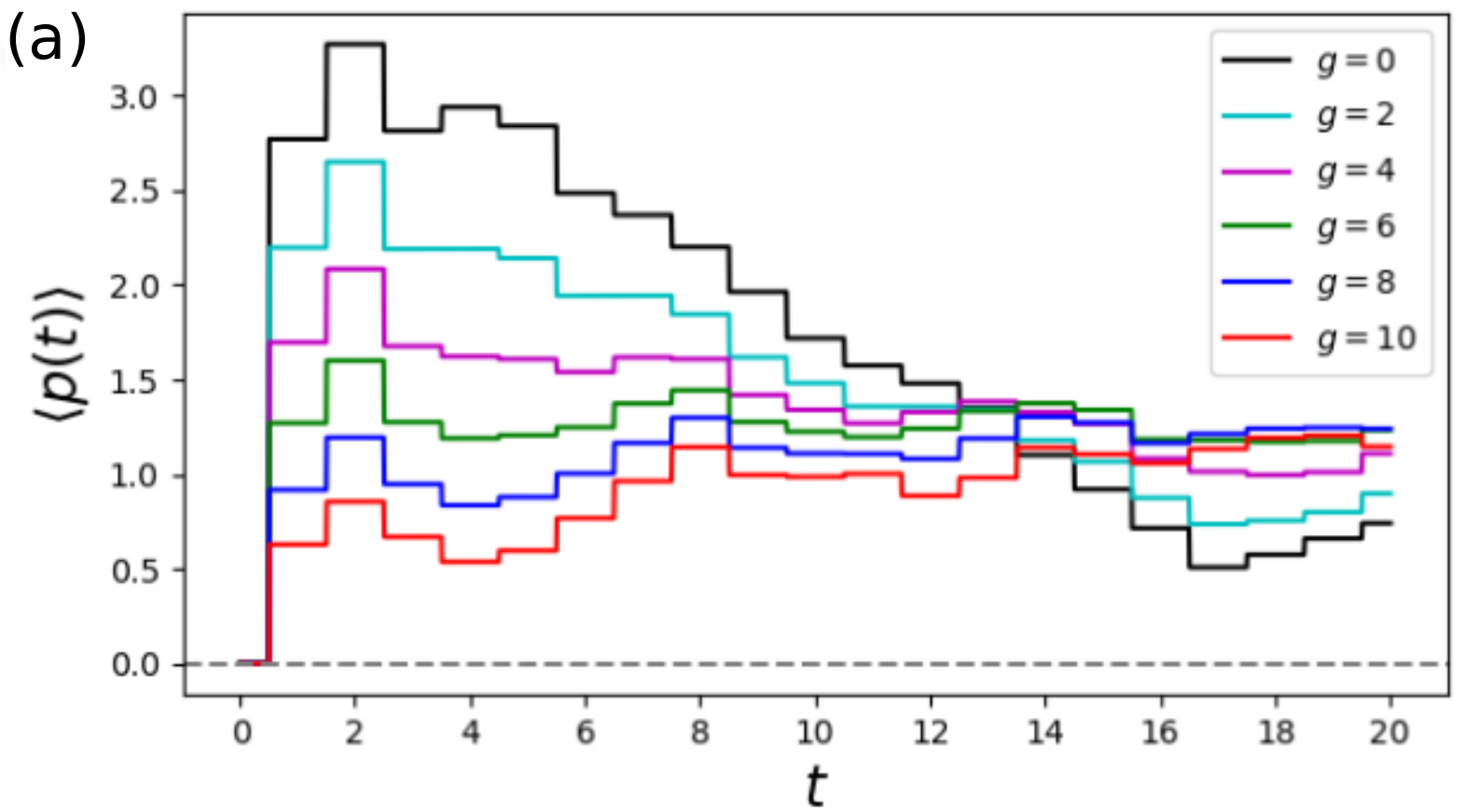}
\includegraphics[width=0.95\columnwidth]{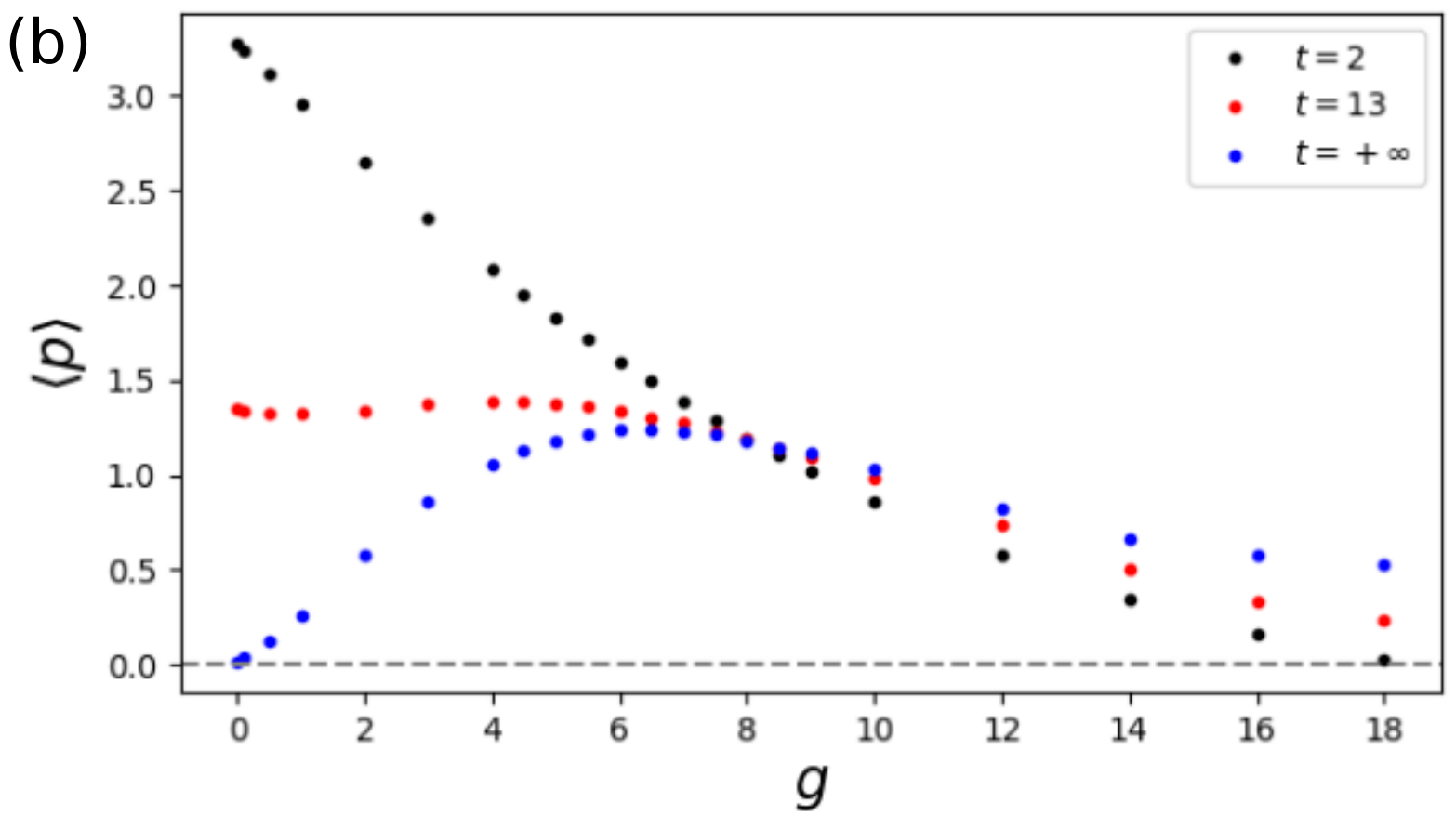}
\caption{\textbf{QBE in the interacting quantum kicked-rotor model}. 
({\it a}) Short-time momentum average for different interaction strengths.
({\it b}) Asymptotic momentum average (blue), momentum at $t=2$ (black), and average momentum at $t=13$ (red). For both plots we set $\alpha=0.5$, $K=5$, $x_0=\pi/2$, $\sigma_k=3$, $\bar{k}=1$, size of the system $L=2\pi \times 512$, discretization in real space $\Delta x=2\, \pi /1024$ and in time $\Delta t= 10^{-2}$.}\label{results1}
\end{figure}

%\subsection{Interacting quantum kicked rotor}
The presence of interactions in some disordered systems may deeply alter the nature of the Anderson transition~\cite{Cherroret2017}.
Interactions can also lead to a breaking of the QBE effect by either destroying localization or $\mathcal{R\,T}$ symmetry. Weakly interacting Bose gases with a contact potential $U(\mathbf{r})=g\, \delta(\mathbf{r})$ are described within a mean-field Hamiltonian density $H= H_0 + g  |\psi(\mathbf{r},t)|^2$, where $H_0$ contains the kinetic and the local disordered potential. The nonlinear term breaks both $\mathcal{T}$ and $\mathcal{R\,T}$ symmetries, leading to the absence of QBE both in momentum and real space. This is in agreement with the analysis in~\cite{Janarek2020Quantum} for the $1$D Gross-Pitaevskii equation (GPE). 
In the following we investigate the QKR model, which presents localization in momentum space and hence displays the QBE in $\langle p(t)\rangle$ in the non-interacting case~\cite{Fishman1982,Tessieri2021Quantum,Sajjad2022}.
In Fig.~\ref{results1} we show the dynamics of $\langle p(t)\rangle$ for the QKR with contact interactions. The mean-field bosonic QKR is governed by the GPE
\beq
i\bar{k} \partial_t \psi  =-\bar{k}^2 \frac{\partial_x^2 \psi}{2}+g|\psi|^2\psi+K\cos{(x)}\sum_{n=-\infty}^{\infty}\delta(t-n-\alpha)\psi.\label{intQKRb}
\eeq
We solve it using third order split-step Fourier method. The initial wave packet is a Gaussian in momentum space with variance $\sigma_k$ and initial ``boost'' $x_0$, $\psi_0(p)=\mathcal{N} \exp(-p^2/2\sigma_k^2-ix_0 p)$.

Interactions are known to destroy dynamical localization in the QKR~\cite{Cao2021}.
Furthermore any finite interaction $g$ breaks \textit{T} symmetry, and hence the \textit{full} QBE is present only for $g=0$. However, $\langle p(t) \rangle$ still displays a \textit{partial} boomerang with a U-turn at $t=2$ for $0<g<g_c\approx 8$. %Moreover, all curves in Fig.~\ref{results1}(a) overlap at $t=13$ for $0<g<g_c$. 
Beyond this critical interaction, there is no signature of the QBE and $\langle p \rangle_{t=2}< \langle p \rangle_{t=\infty}$, where we compute $\langle p \rangle_{t=\infty}$ as an average of $\langle p (t) \rangle$ in the interval $t\in[500,1000]$. This same behavior is observed for other values of $K$ (see the SM~\cite{Supplement} for additional data).

%\section{Conclusion}\label{conclude}
{\it Summary. }
While the QBE was previously found only in \textit{T}-symmetric Hermitian systems with restricted initial conditions,
we showed that QBE can be observed for a wide class of Hamiltonians breaking \textit{T} symmetry and hermiticity, and in a variety of initial states. The QBE is expected to be present in systems of any dimension $d$ and any number $N$ of noninteracting particles. It was shown that sufficient conditions to observe the QBE are \textit{(a)} Anderson localization, \textit{(b)} reality of eigenenergies, \textit{(c)} reflection-time invariance of the ensemble $\{H\}$ of disorder realizations and \textit{(d)} the initial wave function be an eigenstate of the reflection-time operator. 
We observe the breakdown of the QBE when these conditions are not met.
However, these conditions are quite general and hence our results demonstrate the ubiquity of the QBE in localized systems. It is an open question whether these conditions can be further generalized.
%Though our analytical demonstration guarantees the QBE in a very large class of systems, the QBE still appears in situations not included in our derivation, e.g.~in the presence of magnetic field the boomerang appears in the direction orthogonal to the initial momentum, what illustrates the ubiquity of this effect in localized systems. 
We emphasize that the examples discussed in this work have a direct implementation in ultracold systems. Harper-Hofstadter ladders have been realized in e.g.~\cite{Mancini2015} with laser-induced hopping along synthetic dimensions and a complex hopping along the chains producing an effective magnetic field. Local disorder can be added by superimposing an additional incommensurate lattice as in~\cite{Deissler2010} or with a speckle potential~\cite{Billy2008}. 
%Asymmetric hopping amplitudes were proposed for cold atoms in optical lattices via nonlocal jump operator in an effective Lindblad dynamics~\cite{Gong2018,Lindblad1976}.
%Although we focused on a non-Hermitian random hopping model, the QBE holds for a broader class of non-Hermitian systems~\cite{nonHermitian}.
Although in the numerical investigations we focused on Hermitian models, the QBE holds for a broad class of non-Hermitian systems~\cite{nonHermitian}.
Finally, we provided arguments for which mean-field interactions prohibit QBE in bosonic systems. The question whether many-body localized (MBL) phases in interacting systems display QBE remains open~\cite{Zhang2011,Zhang2012}.
Interestingly, Creutz ladders with cross tunnelings can lead to the formation of flat bands and might display disorderless MBL states~\cite{Creutz1999,Zurita2019,Kuno2020,Orito2021}. Also, the presence of momentum-space QBE can be tested in the interacting kicked-rotor model tuning the interaction in a $^7$Li BEC via Feshbach resonances~\cite{Cao2021, Sajjad2022}.

While we were finishing this manuscript we learned about the recent work of Ref.~\cite{Janarek2022}, which has partial overlap with our findings. The authors study a $1$D model with spin-orbit coupling and briefly mention the sufficient conditions to observe the QBE. While our analytical derivation have some similarity with the arguments presented in~\cite{Janarek2022}, our derivation is more general in the sense that we demonstrate the QBE: $(i)$ in non-Hermitian systems with real spectrum, $(ii)$ in a broader class of initial states and $(iii)$ in cases where $H_0$ is not \textit{RT} symmetric if the ensemble $\{H_0\}$ is \textit{RT} invariant. This last point is relevant e.g.~in the model with electric field and in the Hatano-Nelson model~\cite{nonHermitian}. 
%Their analytical arguments have the advantage to guarantee the QBE using either unitary or antiunitary operators, while our analytical demonstration applies to only one of these cases; however, in~\cite{Janarek2022} it is not shown any example for the other case, which seems to prohibit a term like $\textrm{exp}(ik_0x)$ in $\psi_0(x)$. Finally, we present numerical evidence of QBE in several systems, while~\cite{Janarek2022} focuses in a detailed investigation of a single model.

%\begin{acknowledgments}
{\it Acknowledgments. }
We acknowledge L. Tessieri, P. Vignolo, and J. A. S. Louren\c{c}o for useful discussions. We thank N. Cherroret and D. Delande for useful feedback on the manuscript.  
T.M.~ acknowledges CNPq for support through 
Bolsa de produtividade em Pesquisa n.311079/2015-6. 
This work was supported by the Serrapilheira Institute 
(grant number Serra-1812-27802).
We thank the High Performance Computing Center (NPAD) at UFRN for providing computational resources.
%\end{acknowledgments}

%{\it Note added. --}

%\section{Quantum kicked-rotor model}\label{app.qkr}

%\begin{thebibliography}{99}%
\bibliography{Reference}

\begin{thebibliography}{70}
\bibitem [{foo()}]{footeq2}%
  \BibitemOpen
  \href@noop {} {\bibinfo {title} {For each disorder realization $\langle
  x(t)\rangle$ presents fluctuations with time due to the off-diagonal terms of
  {Eq}.~(1) and they oscillate about a time-independent component given by the
  diagonal terms. {Performing} the average over disorder realizations at long
  times causes these time-dependent terms to cancel one another due to the
  random-like nature of the factors that appear in that
  expression.}}\BibitemShut {Stop}%
\end{thebibliography}
%\end{thebibliography}%

\clearpage
\pagebreak
\widetext
\begin{center}
\textbf{\large Supplemental Material:\\[5pt] Ubiquity of the quantum boomerang effect in localized systems\\[10pt]}
\textrm{Flavio Noronha$^{1}$ and Tommaso Macr\`{i}$^{2,1}$\\[3pt]
$^{1}$Departamento de F\'{i}sica Te\'{o}rica e Experimental, Universidade Federal do Rio Grande do Norte, Campus Universit\'{a}rio, Lagoa Nova, Natal-RN 59078-970, Brazil\\
$^{2}$ITAMP, Harvard-Smithsonian Center for Astrophysics, Cambridge, Massachusetts 02138, USA
}
\end{center}
%%%%%%%%%% Merge with supplemental materials %%%%%%%%%%
%%%%%%%%%% Prefix a "S" to all equations, figures, tables and reset the counter %%%%%%%%%%
\setcounter{equation}{0}
\setcounter{figure}{0}
\setcounter{table}{0}
\setcounter{page}{1}
\makeatletter
\renewcommand{\theequation}{S\arabic{equation}}
\renewcommand{\thefigure}{S\arabic{figure}}
\makeatother
%%%%%%%%%% Prefix a "S" to all equations, figures, tables and reset the counter %%%%%%%%%%

\section{Generalization of the analytical derivation}\label{generali}
Some models may fail to meet conditions \textit{(c)} and \textit{(d)} with respect to the operator $\mathcal{RT}$ (see the main text) and still present QBE in real space. One such example is the spin-1/2 interpretation of the Harper-Hofstadter ladder, which was briefly discussed in the main text.
Here we further generalize assumptions \textit{(c)} and \textit{(d)} in order to guarantee the QBE in more general models. Conditions \textit{(a)} and \textit{(b)} are kept the same. Though we use the notation of one-dimensional single-particle models, the considerations below are valid in general contexts.

Suppose that in the system under consideration there exists some (anyone) unitary operator $\mathcal{U}$ that commutes with the position operator $X$ and it is such that \textit{(c)} 
%for each disorder realization $ H$ its \textit{URT} counterpart $\tilde{H}={\mathcal{URT}} H\mathcal{TRU}^\dagger$ is also a disorder realization of the same model, i.e.~
$\mathcal{URT} \{ H\} (\mathcal{URT})^{-1}=\{ H\}$
and \textit{(d)} the initial state is an eigenstate of $\mathcal{URT}$, $\mathcal{URT} |\psi_0 \rangle=\textrm{e}^{i\theta} |\psi_0 \rangle$, $\theta\in \mathds{R}$.
Therefore, using the properties of the operators $\mathcal{U}$, $\mathcal{R}$ and $\mathcal{T}$ and condition \textit{(d)},
\begin{eqnarray}
\langle x(t)\rangle_{ H}
&=&\bra{\psi_0 }  
\textrm{exp}(i H^\dagger t) ~X~ \textrm{exp}(-i H t)
\ket{\psi_0 }\nonumber\\
&=&[\bra{\psi_0 }(\mathcal{URT})^{-1}]  [\mathcal{URT}
\textrm{exp}(i H^\dagger t)(\mathcal{URT})^{-1}] [\mathcal{URT} X (\mathcal{URT})^{-1}]%\nonumber\\ & &\times
[\mathcal{URT} \textrm{exp}(-i H t)(\mathcal{URT})^{-1}]
[\mathcal{URT}\ket{\psi_0 }]\nonumber\\
%
%&=&(\textrm{e}^{-i\theta}\bra{\psi_0 })  [\mathcal{URT} \textrm{exp}(i H^\dagger t)\mathcal{TRU}^\dagger][\mathcal{URT} X\mathcal{TRU}^\dagger] [\mathcal{URT} \textrm{exp}(-i H t)\mathcal{TRU}^\dagger] (\textrm{e}^{i\theta}\ket{\psi_0 })\nonumber\\
%
&=&\textrm{e}^{-i\theta}\bra{\psi_0 }  
\textrm{exp}(-i\tilde{H}^\dagger t) (-X) \textrm{exp}(i\tilde{H} t)
\ket{\psi_0 }\textrm{e}^{i\theta}\nonumber\\
&=& -\langle x(-t)\rangle_{\tilde{H}},\label{QPT2}
\end{eqnarray}
where we have defined $\tilde{H}={\mathcal{URT}} H(\mathcal{URT})^{-1}$. Once $[\mathcal{U},X]=0$ we used $\mathcal{U}X\mathcal{U}^{-1}=X$.
%where we have used condition \textit{(d)} together with the requirements for the operator $\mathcal{U}$. 
Using condition \textit{(c)} we can conclude that $\overline{\langle x(t)\rangle} = -\overline{\langle x(-t)\rangle}$ and, in particular, 
\begin{eqnarray}
\overline{\langle x(+\infty)\rangle} &=& -\overline{\langle x(-\infty)\rangle}.\label{eqder12}
\end{eqnarray}
From Eq.~(3) in the main text and Eq.~(\ref{eqder12})  %$\overline{\langle x\rangle(+\infty)} =\overline{\langle x\rangle(-\infty)}$. 
we have
\begin{eqnarray}
\overline{\langle x(+\infty)\rangle}=0,\label{vanish2}
\end{eqnarray}
which guarantees that the QBE occurs.

The generalization above also applies to the QBE in momentum space. The demonstration is analogous to the ones presented and 
considers only the operator $\mathcal{UT}$ instead of $\mathcal{URT}$, where ${\mathcal U}$ commutes with the momentum operator $P$.

\section{QBE in the QKR}\label{qbeQKR}

The proof of the QBE in momentum space is similar to that in real space. Here we use the noninteracting QKR as example for this demonstration and follow some of the steps presented in~\cite{Sajjad2022}.
In general it is required that {\it (a)} the Hamiltonian presents dynamical localization, {\it (b)} all eigenenergies are real, \textit{(c)} $\mathcal{T}\{H\}\mathcal{T}^{-1}= \{H\}$ and \textit{(d)} $\mathcal{T} |\psi_0 \rangle=\pm |\psi_0 \rangle$. Condition \textit{(c)} is trivially met in the QKR with $\alpha=1/2$ once the model becomes \textit{T} symmetric. In this case we can consider that the ensemble $\{H\}$ is composed by only one Hamiltonian $H$. The QKR does not require disorder to present dynamical localization; instead, the distribution of quasimomenta in $\ket{\psi_0}$ plays the role of pseudo disorder. Therefore there is no need to define the average $\overline{(\cdots)}$ in the QKR.

\begin{figure}[ht]
\centering
\includegraphics[width=0.5\columnwidth]{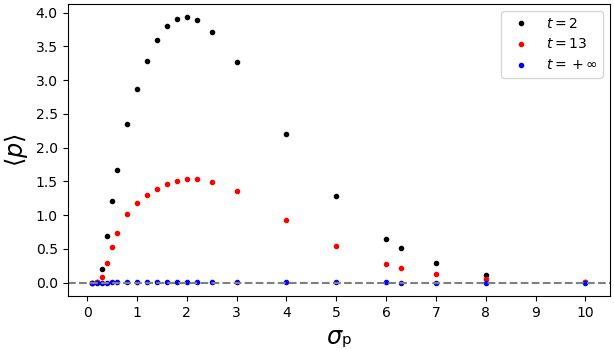}
\caption{\textbf{Dependence of QBE on the initial width of the wave packet}. 
We propagate the wave function in the non-interacting QKR using Floquet operators as in~\cite{Tessieri2021Quantum} and used $n=1000$ values of integer momenta, $n_d=1000$ values of quasi-momentum $\beta$, $x_0=\pi/2$, $K=5$ and several values of $\sigma_p$.}\label{qkr1}
\end{figure}

\begin{figure}[b]
\includegraphics[width=0.49\columnwidth]{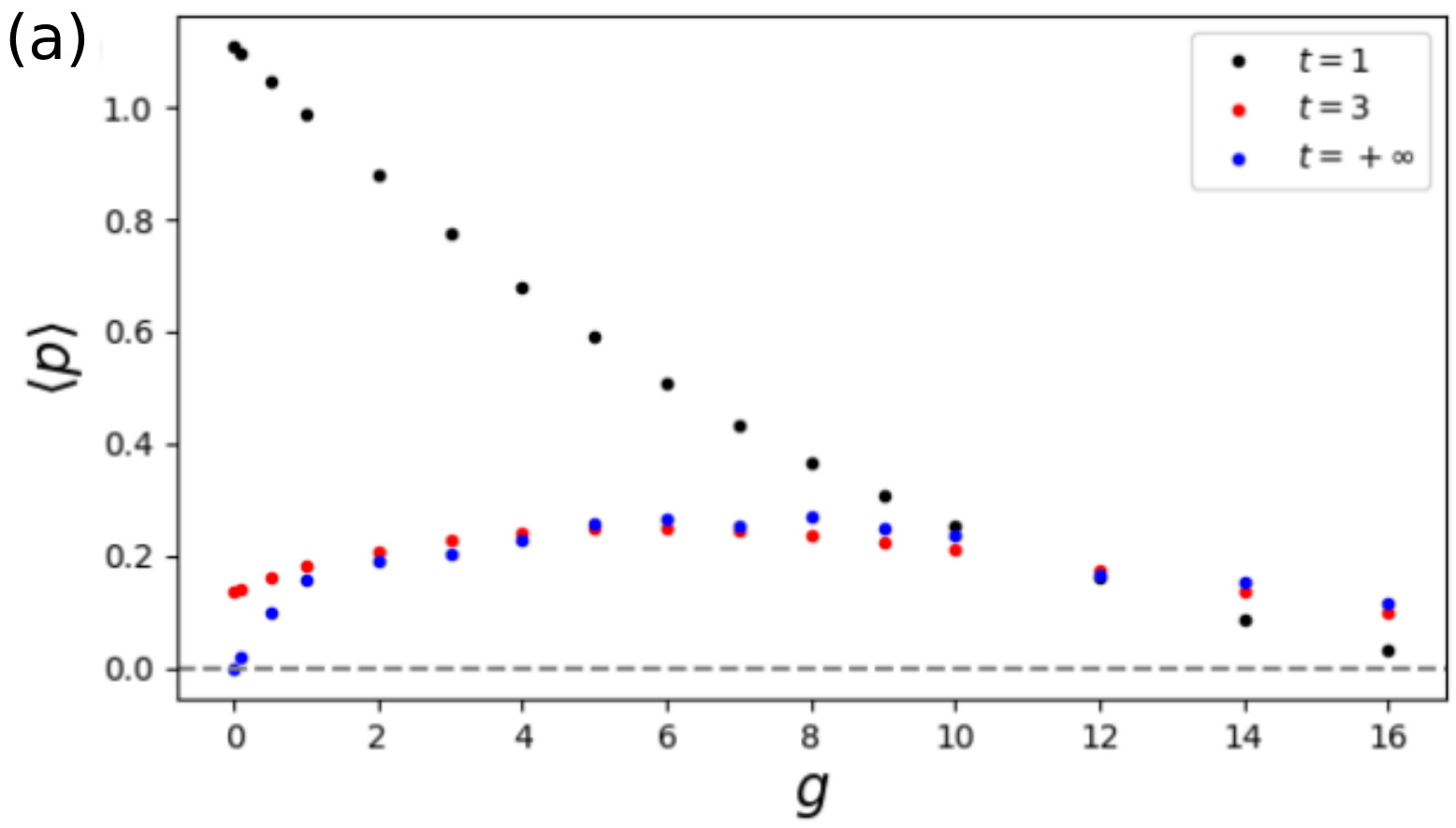}
\includegraphics[width=0.49\columnwidth]{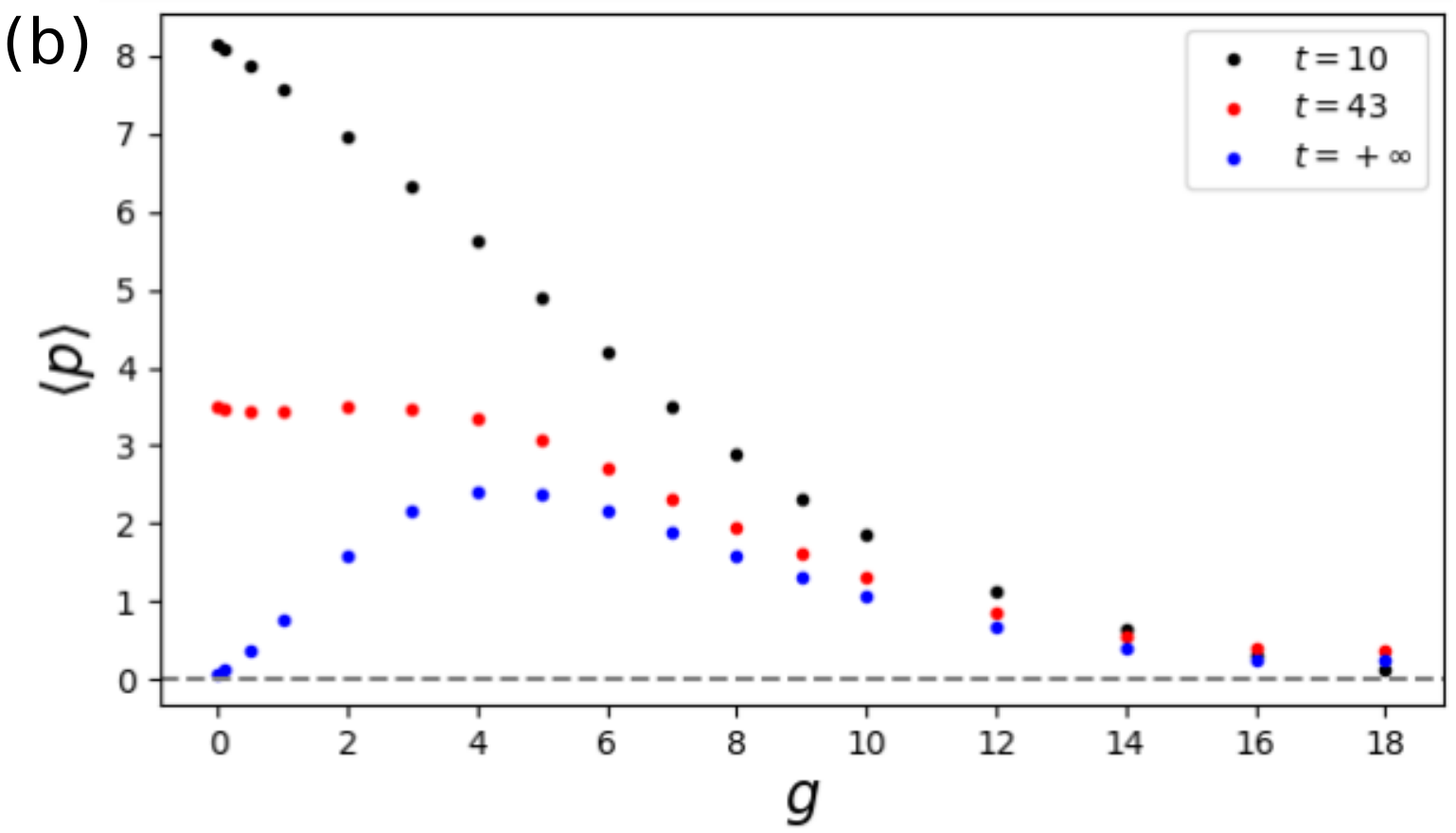}
\caption{\textbf{Partial QBE in the interacting quantum kicked-rotor model}. 
In ({\it a}) we use $K=2$ and show the average momentum at $t=1$ (black) and at $t=3$ (red). 
In ({\it b}) we use $K=8$ and show the average momentum at $t=10$ (black) and at $t=43$ (red). 
The asymptotic momentum average $\langle p \rangle_{t=\infty}$ is shown in blue and is computed as an average of $\langle p (t) \rangle$ in the interval $t\in[500,1000]$.
For both plots we set $\alpha=0.5$, $x_0=\pi/2$, $\sigma_k=3$, $\bar{k}=1$, size of the system $L=2\pi \times 512$, discretization in real space $\Delta x=2\, \pi /1024$ and in time $\Delta t= 10^{-2}$.
}\label{results22}
\end{figure}

In the case of the QKR, which has a time-dependent Hamiltonian, we can expand $\ket{\psi_0}=\sum_n c_n \ket{\phi_n}$ in terms of the eigenvectors of its Floquet operator, $U\ket{\phi_n}=\exp(-i\epsilon_n) \ket{\phi_n}$. The eigenstates of $U$ and $U^\dagger$ are identical, and correspond to the eigestates of the same self-adjoint,
time-independent Floquet Hamiltonian.
At any time $t=l\in \mathds{N}$, the average momentum is given by %$\langle x(t)\rangle = \int_{-\infty}^{+\infty}\, x \, \left|\psi (x,t)\right|^2\, dx$, 
\beq 
\langle p(t=l)\rangle=\sum_{n,m} c_n c_m^*\,\textrm{exp}[-i(\epsilon_n - \epsilon_m)l]\bra{\phi_m}P \ket{\phi_n}, \label{expect2}
\eeq 
where $P$ is the momentum operator. Note that this same expression is expected to hold for other models with a time-independent (Floquet) Hamiltonian that meets condition~\textit{(b)}. Condition \textit{(a)} leads to the diagonal ensemble 
\beq 
\langle p(t=+\infty)\rangle=\sum_{n} |c_n |^2\,\bra{\phi_n}P \ket{\phi_n}. \label{expect22}
\eeq 
We stress here that in the QKR the role of disorder realizations is played by the quasimomentum distribution of $\ket{\psi_0}$, which leads to a sufficiently large set of non-zero numbers $\{c_n\}$ that guarantee that the off-diagonal terms cancel out. Other models may require an average $\overline{(\cdots)}$ over a large set $\{H\}$ to ensure the diagonal ensemble.
A similar expression holds for $t\to-\infty$ and then
%
%Equation~(\ref{expect}) in the main text can be rewritten replacing $x$ with $p$,
%\begin{eqnarray}\nonumber
%\left|\psi (p,t)\right|^2 &=& \bra{\psi_{0} }\textrm{exp}(+i H^\dagger t)\ketbra{p}{p}\textrm{exp}(-i Ht)\ket{\psi_0 }\\
%&=& \sum_{n,m} \phi_{m}(p)^{*} \phi_{n}(p)\,\textrm{exp}[-i(\epsilon_n - \epsilon_m^*)t]  \braket{\psi_{0} }{\phi_m}\braket{\phi_n}{\psi_{0} }.\label{psipt}
%\end{eqnarray}
%
%Due to dynamical localization one finds at $t\to \pm\infty$ the diagonal ensemble Eq.~(\ref{diagems}) in momentum space,
%\begin{eqnarray}
%\overline{\left|\psi (p,\pm\infty)\right|^2}&=&\sum_n\overline{\left|\phi_{n}(p)\right|^2\left|\braket{\psi_0 }{\phi_n}\right|^2}.\label{diagens2}
%\end{eqnarray}
%Notice that in the case of the QKR the average $\overline{(\cdots)}$ is done considering the quasimomenta of the initial state, which plays the role of disorder in this model~\cite{Fishman1982,Tessieri2021Quantum}.
%
%Therefore %$\overline{\langle p(+\infty)\rangle} = \overline{\langle p(-\infty)\rangle}$.
\begin{eqnarray}
\langle p(+\infty)\rangle &=& \langle p(-\infty)\rangle.\label{eqdiagp22}
\end{eqnarray}

Following Eq.~(4) in the main text, considering $\mathcal{T}$ instead of $\mathcal{RT}$, the momentum operator $P$ instead of $X$ and defining $\tilde{H}=\mathcal{T} H \mathcal{T}^{-1}$ we find $\langle p(t)\rangle_{ H}
= -\langle p(-t)\rangle_{\tilde{H}}$ using condition~\textit{(d)}. %using $\mathcal{QT} P\, \mathcal{TQ}=-P$,
%\begin{eqnarray}
%\langle p(t)\rangle_{ H} = -\langle p(-t)\rangle_{\tilde{H}},\label{eqpmp}
%\end{eqnarray}
Therefore, using \textit{(c)} we find %$\overline{\langle p(+\infty)\rangle} = -\overline{\langle p(-\infty)\rangle}$.
\begin{eqnarray}
\langle p(+\infty)\rangle &=& -\langle p(-\infty)\rangle.\label{eqder2}
\end{eqnarray}
From the above relations we have $\langle p(+\infty)\rangle=0$,
%\begin{eqnarray}
%\overline{\langle p(+\infty)\rangle}=0,\label{vanishp}
%\end{eqnarray}
which guarantees that the QBE occurs in the QKR.

The above derivation is more general than the one presented in~\cite{Sajjad2022} in the sense that here we do not use that the Hamiltonian is parity symmetric; we just use condition \textit{(c)}.
We emphasize that the above derivation may remain valid for other models with dynamical localization when considering the average $\overline{(\cdots)}$.

In the QKR the amplitude of the boomerang depends on the width of the initial wave packet. See Fig.~\ref{qkr1} for the non-interacting case.

In Fig.~\ref{results22} we display additional data for the interacting QKR of Eq.~(10) in the main text. Using $K=2$ and $K=8$ we find evidence for the existence of a critical interaction $g_c$ below which there exists a partial boomerang with a U-turn.

\section{Additional data for the Harper-Hofstadter ladder model}\label{Creutzadd}

\begin{figure}[h]
\centering
\includegraphics[width=\columnwidth]{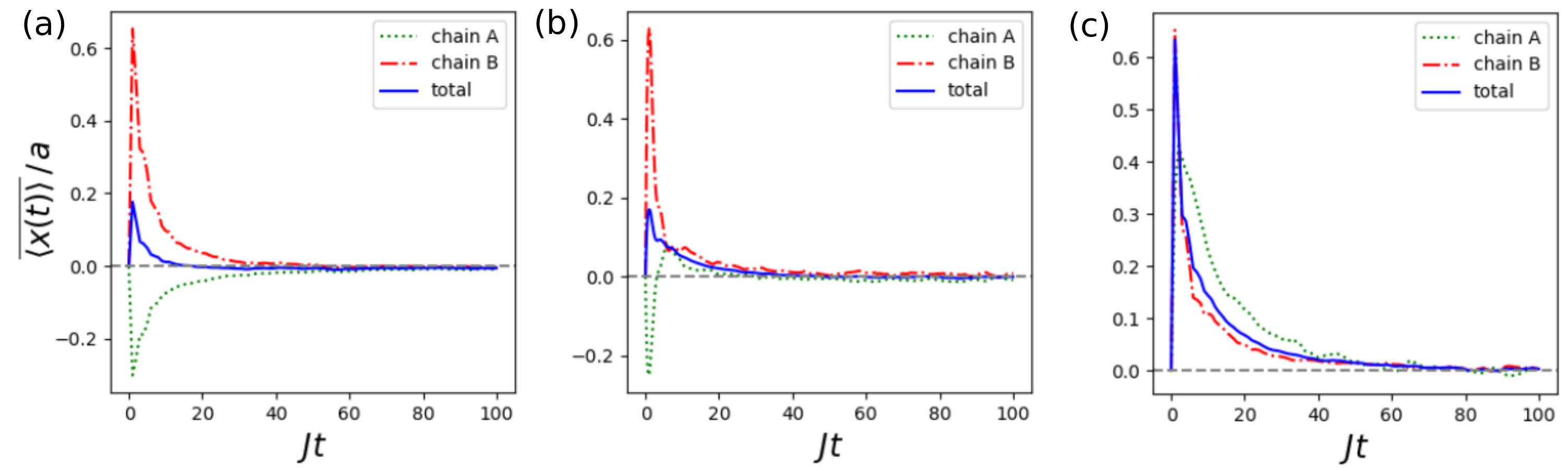}
\caption{\textbf{Harper-Hofstadter ladder model}. 
We show the disorder averaged center of mass in chain A (green dotted line), chain B (red dash-dotted line) and averaged in the whole system (blue solid line). We used $W/J=6$, $N=2\times 10^2$, $n_d=5\times10^4$, $\sigma/a=10$, $k_0a=0.2$ and $\phi=(\pi/2)(\sqrt{5}-1)/2$. We show (a) two disconnected chains with $\Omega/J=0$ and initial state $\psi_{0+}$, (b)~two connected chains with $\Omega/J=0.3$ and initial state $\psi_{0+}$, and (c) two connected chains with $\Omega/J=0.3$ and initial state in chain B only, $(\psi_{0+}-\psi_{0-})/\sqrt{2}$. }\label{figcreutz2}
\end{figure}

If the two chains in the Harper-Hofstadter ladder model are disconnected ($\Omega=0$), it is possible to use values for the magnetic flux $\phi$ and momentum $k_0$ such that the particle propagates in opposite directions in each chain [see Fig.~\ref{figcreutz2}(a)]. Using a finite $\Omega$ leads to a tendency for the particle to propagate in the same direction in both the chains [see Fig.~\ref{figcreutz2}(b)].
When the initial wave packet is nonzero in only one of the chains, a finite inter-lattice hopping $\Omega$ allows the particle to partially migrate to the other chain [see Fig.~\ref{figcreutz2}(c)]. In all these cases the QBE is present in each chain.

\end{document}